\documentclass[12pt]{iopart}

\usepackage{iopams}  
\usepackage{esint}

\begin{document}

\eqnobysec
\title[New example of integrable nonlinear coupled equations...]{New example of integrable nonlinear coupled equations with exact asymptotic singular solution in the context of laser-plasma interaction}

\author{A. Latifi}

\address{Department of physics, Faculty of sciences, Qom University of Technology, Qom, Iran}
\ead{latifi@qut.ac.ir}
\vspace{10pt}
\begin{indented}
\item[]August 2015
\end{indented}

\begin{abstract}
A new set of nonlinear coupled equations is derived in the context of small amplitude limit of the general wave equations in a fluid type warm electrons/cold ions plasma irradiated by a continuous laser beam. This limit is proved to be integrable by means of the spectral transform theory with singular dispersion relation. An exact asymptotic solution is obtained. This model accounts for a nonlinear mode coupling of the electrostatic waves with the ion sound wave, and is shown to be unstable and does not propagate any stable small amplitude solution. This instability is understood as a continuous secular transfer of energy from the electrostatic wave to the ion sound wave through the ponderomotive force. The exact mechanism of this transfer is exposed. The dynamics of this energy transfer results in a singular asymptotic behavior of the ion sound wave which explains the low penetration of the incident laser beam.
\end{abstract}

\vspace{2pt}


\submitto{\NL}

\maketitle

\section{Introduction}

\hskip9mm Recent progress in laser generation has aroused a great interest in the problem of laser-plasma interaction, which is the source of various nonlinear effects such as self-focusing, self-modulation, filamentation and several parametric instabilities \cite{bmp,kd,shir,bgks,qtx}. These phenomena have applications in many advanced physical events such as inertial confinement fusion, higher order harmonic generation, X-rays sources and particle acceleration \cite{pk,bssm,lee,eskt}.\\

This work regards the general theory of nonlinear coupled waves in plasmas, in the context of a two-components non magnetized fluid-type plasma. In a plasma produced and irradiated by laser beam, the electrostatic wave (EW) induces scattered waves by means of different types of stimulated emission \cite{kt,td}. We are interested here in the process of stimulated Brillouin (back) scattering (SBS) for longitudinal EW. The resulting incident and reflected high frequency EW can drive some low frequency waves of acoustic type, such as the Ion Sound Wave (ISW), by means of the ponderomotive effect on the local charge density. In the present paper, it is shown here that this nonlinear effect is at the origin of plasma instability.\\

It is well known that the model of nonlinear propagation of ISW, when the ponderomotive effects are neglected (no mode-coupling), is the famous Kortoveg de Vries (KdV) equation  \cite{kt,td}. This equation has the advantage of being integrable, which allows to describe the solution on the basis of its "nonlinear Fourier spectrum" (see f.i. \cite{cd} and \cite{ks}) . When the ISW is decomposed into his harmonic modes, the KdV equation maps into the nonlinear Schr\"odinger equation (NLS), an other integrable model  \cite{kt,td,cd,ks}. Both models propagate soliton solutions which provide the general fluid equations with stable small amplitude localized coherent structures. But these models ignore the possibility of energy transfer from the long wavelength domain to the dissipative  short wavelength range.\\

This energy transfer is considered by including in the momentum transfer equation for the electrons, the poderomotive force (low frequency effect of the high frequency electrostatic field) which results on the nonlinear mode coupling. In this approach Karpman  \cite{karpman} derived an integrable model, which however does not account for dispersion or nonlinear dynamical effects (ISW nonlinearity). Still this model has proved to be very useful and has been slightly expanded in \cite{yo} and \cite{kaup}. All these models and their stability properties are discussed in \cite{ls}.\\

Later, M. Casanova and al. \cite{clp}, have shown that the wave dynamics in plasma is well described by a model where the ISW is nonlinear and is coupled to the electrostatic field through the poderomotive term. In particular, the ISW nonlinearity has been shown to play a crucial role in the Brillouin reflexivity.\\

In this paper, starting from the most general one-dimensional fluid equation for a plasma with a Maxwellian velocity distribution \cite{kt}, a model for the nonlinear propagation of the ISW coupled to the EW (the pump wave and the wave reflected through SBS) is established. This model is obtained as an exact asymptotic limit in the regime of "warm electrons-cold ions" and small amplitude of the pump wave. The technique used is the well known multisclae expansion \cite{zk,sandri}, and the model exposed in this paper, appears then as universal in the sens given in \cite{calo}.This model is a new nonlinear evolution equation with unexpected properties. The integrability permits to obtain general theorems about the nature and the behavior of the solution. In particular,  it is shown that the EW transfers its energy to the ISW through a highly nonlinear and unstable process : any infinitely small initial disturbance of the ISW asymptotically evolves towards a singular (soliton-like) solution.

\normalsize
\section{Derivation}

\hskip9mm Let consider a linearly polarized electrostatic field vector $\vec E = \hat{z}  E (z,t)$ with a frequency close to plasma frequency $\omega_0 = 4\pi  e^2 n_0 / m_e$ with some broadening represented by the following wave paquet 
\begin{equation}
E(z,t)=\int_{-\infty}^{+\infty}d\omega \tilde{ E}  (\omega ,z,\tau)\mathrm{e}^{-\rmi\omega t}\hskip1mm.
\end{equation}
Note that $m_0$ and $-e$ are respectively the mass and the charge of electrons and $n_0$ is the ambient ion density. $ \tilde{ E}  \left( \omega ,z,\tau\right )$ is the slowly varying envelop of $E(z,t)$ where $\tau$ is the slow scaled time defined below \ref{slowvar}.\\

The low frequency effect on the electrons due to the high frequency field $E$, results in the ponderomotive force obtained by considering the real part of $[\delta z (\partial E / \partial z)]$ around the average position $<z>$ where  $\delta z$ is the complex valued displacement and solution of
\begin{equation}
m_e (\delta z)_{tt }= e E\hskip1mm.
\end{equation}
The ponderomotive force reads then
\begin{equation}
f_p=-\frac{e^{2}}{2m_e}\hskip1mm\partial z\int_{-\infty}^{+\infty}d\omega \hskip1mm {\omega}^{-2} {\left |\tilde{ E}  (\omega ,z,\tau)\right |}^2\hskip1mm.
\end{equation}
Using the ratio of electron and ion masses $\epsilon_1\doteqdot(m_e$/$m_i)^2$ and the electron Debeye wavelength $\lambda_D^2=K_BT_e$/$4\pi n_0 e^2$ (with $T_e$ the electron temperature and $K_B$ the Boltzman constant), a dimensionless set of variables can be set as follows :
\begin{equation}
z' = \lambda_D^{-1} z\qquad ,\qquad t' = \epsilon_1\omega_0t\hskip1mm.\label{dimlessvar}
\end{equation}
in these variables, we have the usual system of fluid-type equations  \cite{kt} with the only additional term $f_p$ in the momentum transfer equation for the electrons   \cite{lelat}.\\

Moreover, the Maxwell equation for the EW is obtained on the basis of the dispersion relation \cite{kt}
\begin{equation}
\omega^2=\omega_p^2+3V_{Te}^2k^2  \label{displaw}
\end{equation}
where $V_{Te}=\lambda_D\omega_0$, is the thermal electron velocity, $k$ is the wave number and $\omega_p$ stands for the plasma frequency
\begin{equation}
\omega_p^2=\omega_0^2(1+q_e)\hskip1mm.
\end{equation}
$q_e$ is the fractional change in the electron density of average value $n_0$, namely
\begin{equation}
n_e=n_0\left[1+q_e(z,t)\right]\hskip1mm. 
\end{equation}
Now, redefining the fields as
\begin{equation}
\tilde{E}'=\tilde{E}\left(\frac{e^2}{2m_e\omega^2K_BT_e}\right) \hskip4mm ,\hskip4mm \phi'=\frac{e}{K_BT_e}\phi\hskip4mm ,\hskip4mm v_i'=v_i\left(\frac{m_i}{K_BT_e}\right)^{1/2}
\end{equation}
where $\phi$ is the electrostatic potential and $v_i$ the ion velocity, the momentum transfer equation for electrons reads  \cite{kt,lelat}
\begin{equation}
\frac{\partial \phi'}{\partial z'}-\frac{1}{(1+q_e)}\frac{\partial q_e}{\partial z'}-\frac{\partial}{\partial z'}\int_{-\infty}^{+\infty}d\omega \hskip1mm {\left |\tilde{ E}'  (\omega ,z',\tau)\right |}^2=0 \hskip1mm ,\label{momtr}
\end{equation}
and for the ions 
\begin{equation}
\frac{\partial v_i'}{\partial t'}+v_i'\hskip1mm\frac{\partial v_i'}{\partial z'}=-\frac{\partial \phi'}{\partial z'}\hskip1mm.
\end{equation}
The above system has to be completed with the continuity equation
\begin{equation}
\frac{\partial q_i}{\partial t'}+\frac{\partial}{\partial z'}\left[ (1+q_i)\hskip1mm v_i' \right]=0\hskip1mm , \label{continuity}
\end{equation}
and the Poisson equation
\begin{equation}
\frac{\partial ^2 \phi'}{\partial {z'}^2}= q_e - q_i \hskip1mm ,\label{poisson}
\end{equation}
where $q_i$ in \ref{continuity} and \ref{poisson} is the fractional change in the ion density $n_i=n_e(1+q_i)$.\\

Let define a slowly varying set of variables in the comoving frame of the ISW at the sound speed $C_s =(K_B T_e$/$m_i)^{1/2} =\epsilon_1\omega_0\lambda_D$ ;
\begin{equation}
\xi = \epsilon_1(z'-t') \qquad ,\qquad \tau ={( \epsilon_1)}^{3/2}t\hskip1mm. \label{slowvar}
\end{equation}
Now, by expanding the electrostatic field $\phi$, the density variations $q_e$ and $q_i$ and the scaled ion velocity $v_i'$, in powers of $\epsilon$ as
\begin{equation}
v_i'=\epsilon_1 v_i^{(1)}+\epsilon_1^2 v_i^{(2)}+O(\epsilon_1^3)  \label{expan}
\end{equation}
and by defining $\mathcal{E}$ as follows
\begin{equation}
\tilde E'={(\epsilon_1)}^{1/2}\left [\mathcal{E} + O(\epsilon_1)\right]\hskip1mm, \label{EWexpan}
\end{equation}
the set of equations \ref{momtr}-\ref{poisson}, after integration of \ref{momtr}, gives at the first order
\begin{equation}
r\doteqdot q_e^{(1)}= q_i^{(1)}=\phi^{(1)}=v_i^{(1)} \label{defr}
\end{equation}
and at the second order, by eliminating $q_i^{(2)}$ , $q_e^{(2)}$ , $v_i^{(2)}$ and $\phi^{(2)}$ , one obtains the following evolution for $r$ (defined in \ref{defr}):
\begin{equation}
2r_\tau+2rr_\xi+r_{\xi\xi\xi}=-\partial_\xi\int_{-\infty}^{+\infty}3\omega_0\nu d\nu \hskip1mm {\left |\mathcal{E}\right |}^2\hskip1mm .\label{firsteq}
\end{equation}
On the other hand, the Maxwell equation from \ref{displaw} reads
\begin{equation}
\left[\frac{\partial^2}{\partial t^2}-3{(V_{Te})}^2\frac{\partial^2}{\partial z^2}\right]E(z,t)=-\omega_0^2\left[ 1+q_e(z,t)\right ] E(z,t)\hskip1mm. \label{max}
\end{equation}
Using the change of variable \ref{dimlessvar}, the scalings \ref{slowvar}, \ref{EWexpan} and the expansion \ref{expan}, the equation \ref{max} gives at first order
\begin{equation}
\partial_{\xi\xi}\mathcal{E}+(\nu^2-\frac{r}{3})\mathcal{E}=0 \hskip1mm ,\label{secondeq}
\end{equation}
where the parameter $\nu$ is defined through the relative difference between the squared frequency $\omega$ of the applied laser beam and the squared plasma frequency $\omega_0$ :
\begin{equation}
\frac{\omega^2-\omega_0^2}{\omega_0^2}=3\hskip1mm \epsilon_1\nu^2\hskip1mm.
\end{equation}
The system \ref{firsteq}-\ref{secondeq} representing the nonlinear propagation of the EW coupled to the ISW does not account for SBS and has been shown to be not integrable \cite{lelat}. To take SBS into account, one has to allow the electric field envelop $\mathcal{E}$ to contain simultaneously at first order, a left-going wave $\exp\left[ i(\omega_1\tau+k_1)\right]$ and a right-going wave $\exp\left[ i(\omega_2\tau+k_2)\right]$ with small slowly varying amplitudes $\epsilon_2a_1(\xi,\tau)$  and  $\epsilon_2a_2(\xi,\tau)$ respectively, $\epsilon_2$ being a small parameter submitted to the constraint \ref{smallvar2}. The wave numbers moduli are closed to a same average value $k_e$, namely :
\begin{equation}
k_1=k_e+O(\epsilon_2) \qquad,\qquad k_2=-\hskip1mm k_e +O(\epsilon_2)
\end{equation}
which means
\begin{equation}
\mathcal{E}(\xi,t)=\epsilon_2\left[a_1\mathrm{e}^{\displaystyle{\rmi(\omega_1\tau+k_1z)}}+a_2\hskip1mm\mathrm{e}^{\displaystyle{\rmi(\omega_2\tau+k_2z)}}\right]\label{leftright}
\end{equation}
and the frequencies are related through (resonant scattering conditions)
\begin{equation}
\omega_1=\omega_2+\omega_s \qquad,\qquad k_1= k_2 +k_s
\end{equation}
where $\omega_s$ and $k_s$ stand for the ISW parameters. In the frame $(\xi,\tau)$, the dispersion relation reads $\omega_s=\frac{1}{2}{(k_s)}^3$\hskip1mm.\\

In the laboratory frame $(X,T)$ with the scaled variables \cite{zk} defined as follows
\begin{equation}
X=\epsilon_2\left[\xi+\frac{3}{2}{(k_s)}^2\hskip1mm\tau\right]\qquad,\qquad T=3\hskip1mm {(\epsilon_2)}^2\hskip1mm k_s\hskip1mm \tau\hskip1mm ,
\end{equation}
we consider $r(\xi,\tau)$ (defined in \ref{defr}) to be a superposition of its harmonic modes expanded in powers of $\epsilon_2$ as the formal series
\begin{equation}
r={(\epsilon_2)}^2\hskip1mm\tilde{q}_0+\sum_{n=1}^{N}{(\epsilon_2)}^n\hskip1mm\tilde{q}_n\hskip1mm\mathrm{e}^{\displaystyle{\rmi  n(\omega_s\tau+k_2\xi)}}+C.C.\label{harm1}
\end{equation}
with
\begin{equation}
\tilde{q}_n=\sum_{j}^{M}q_{nj}(X,T){(\epsilon_2)}^j+O(\epsilon^{M+1})\qquad,\qquad n=0,...,N\hskip1mm.\label{harm2}
\end{equation}
For $N=2$ and $M\leqslant3$, by means of the standard multiscale techniques\cite{zk,calo,lelat} i.e. by inserting \ref{leftright} and \ref{harm1} into \ref{firsteq} and \ref{secondeq}, then examining the relevant orders in the expansions in powers of $\epsilon_2$ and averaging over the fast oscillation $\exp(i\omega_1\tau)$ (which means : look at the coefficients of $\exp(i\omega_s\tau)$ only, known as the "rotationg wave approximation"), one obtains for
\begin{equation}
q=\frac{1}{3\rmi k_s}q_0(X,T)\qquad,\qquad \epsilon_2\hskip1mm\zeta=\frac{k_e^2-\nu^2}{2k_e}
\end{equation}
the following system of coupled equations
\begin{equation}
\cases{
-\rmi q_{_T}+\frac{1}{2}q_{_{XX}}-q\left|q\right|^2=\rmi\displaystyle\frac{\omega_0}{12}\displaystyle\int_{-\infty}^{\infty}d\zeta a_1\overline{a}_2\\
{a_1}_{,X}+\rmi\zeta a_1=qa_2\\
{a_2}_{,X}-\rmi\zeta a_2 =\overline{q}a_1
}\hskip3mm.\label{system}
\end{equation}
the left going wave of amplitude $a_1(X,T)$ results from the incident (pump) laser beam with wave number $\zeta$ and initial given amplitude, say $A(\zeta)$. The right going wave of amplitude  $a_2(X,T)$ is the wave reflected by SBS.\\

Note that the two multiscale analysis with the small parameters $\epsilon_1$ and $\epsilon_2$ applied in this section have to be consistent, which is guaranteed by the ordering
\begin{equation}
\epsilon_1\ll\epsilon_2^2\ll\epsilon_2\ll 1\hskip1mm.\label{smallvar2}
\end{equation}
 To the system \ref{system}, we adjoint the following initial/boundary values
\begin{equation}
\cases{
q(X,0)\in L^1(\mathbb{R})\\
\displaystyle\lim_{X\to +\infty}a_1(\zeta , X,T)= A(\zeta) \mathrm{e}^{-i\zeta X}\\
\displaystyle\lim_{X\to -\infty}a_2(\zeta , X,T)=0
}\hskip3mm.\label{cond}
\end{equation}
Thus, the system of coupled equations \ref{system} completed by the initial/boudary conditions \ref{cond} is the central point of discussion in the following sections.

\section{Method of resolution}

The system \ref{system} with the initial/boundary conditions \ref{cond} belongs to a general class of integrable nonlinear evolution \cite{leon} with a singular dispersion law \cite{leon,kp,leon2,leon3}. This system is a special case of the following system for the $2 \times 2$ matrices $\mu (k,x,t)$ and $Q(x,t)$ :
\begin{equation}
\cases{
Q_t-\frac {\rmi}{2} \sigma_3 Q_{xx}+\rmi \sigma_3 Q^3 =\rmi \left [ \sigma_3 , \langle \mu \sigma_3 \mu^{-1} \rangle \right ]\\
\mu_x +\rmi k \left [ \sigma_3 , \mu \right ] = Q \mu
}\hskip3mm\displaystyle.\label{evolsys}
\end{equation}
where 
\begin{equation}
Q =\pmatrix {0 & q \cr  \bar{q} & 0 } \hskip5mm , \hskip5mm \mu = \pmatrix{  \mu_{11} &  \mu_{12} \cr  \mu_{21} & \mu_{22} }  \hskip5mm , \hskip5mm \sigma_3 = \pmatrix{ 1 & 0 \cr   0 & -1 }
\end{equation}
and 
\begin{equation}
\langle \mu \sigma_3 \mu^{-1} \rangle \doteqdot \frac{1}{2\pi i}\iint_\mathbb{C} d\lambda \wedge d\bar{\lambda}\hskip1mm g(\lambda , t) \mu (\lambda , x,t) \hskip1mm \sigma_3\hskip1mm\mu ^{-1} (\lambda ,x,t)\hskip1mm ,
\end{equation}
for any given distribution $g(k,t)$ in the complex plane. We have also
\begin{equation}
d\lambda \wedge d\bar{\lambda} \doteqdot -2\rmi \hskip1mm d\lambda_R\hskip1mm d\lambda_I \qquad \textrm{for} \qquad \lambda = \lambda_R + \rmi\lambda_I
\end{equation}
The above system has been shown to be integrable \cite{kp,leon2,leon3}. It is related to the general theory of the spectral transform \cite{cd,ks} in the context of the singular dispersion relation \cite{lelat}. A complete proof and details are given in the Appendix.\\

One can summarize the method of resolution by the following set of linear steps :\\

{\bf Step 1 :} Given $q(x,0)$ on $ L^1(\mathbb{R})$, solve the following Volterra equations for the vector $\mu^+(\zeta , X, 0)$ and $\mu^-(\zeta , X, 0)$ with components $\mu_1$ and $\mu_2$
\begin{equation}
\cases{
\mu_1^+=1-\displaystyle \int_{X}^{+\infty}dX'q\mu_2^+\\
\nu_2^+=1-\displaystyle \int_{-\infty}^XdX'q\nu_1^+
} \qquad ; \qquad
\cases{
\mu_1^-=1-\displaystyle \int_X^{+\infty}dX'q\mu_2^-\\
\nu_2^-=1-\displaystyle \int_X^{+\infty}dX'q\nu_1^-
} \label{volterra}
\end{equation}
with $\displaystyle{\nu=\mu\exp(2\rmi\zeta X)}$\hskip1mm .\\

The vector $\mu^+$ (respectively $\mu^-$) is homomorphic in $\Im (\zeta) > 0$ (respetively in $\Im (\zeta) < 0$) and the discontinuity $\mu^+-\mu^- $, as $\zeta$ crosses the real axis leads to the following Riemann-Hilbert problem for $\zeta \in \mathbb{R}$
\begin{equation}
\mu^+(\zeta+\rmi 0)-\mu^-(\zeta-\rmi 0) = R(\zeta+\rmi 0)\hskip1mm \mathrm{e}^{2\rmi\zeta X} \sigma_1\overline{\mu^-}(\zeta -\rmi 0)\label{riemhil}\hskip1mm .
\end{equation}
The function $R(\zeta +\rmi 0)$ is called the nonlinear Fourier Transform of $q(X,0)$ and $\sigma_1$ is the usual Pauli matrix. $R$ is computed out of the Volterra equation \ref{volterra} for given $q(X,0)$ by
\begin{equation}
R(\zeta , 0)=\int_{-\infty}^{+\infty} dX'\bar{q}(X',0)\mu_1^+(\zeta , X',0)\mathrm{e}^{-2i\zeta X'}\hskip1mm .\label{volterraR}
\end{equation}
\vskip2mm
\hskip9mm This first step solves the direct problem, namely the determination of $R(\zeta)$  from $q(X,0)$.\\

{\bf Step 2 :} When $q(X,T)$ obeys the nonlinear evolution equation of \ref{system}, the time dependance of $R$ is given by
\begin{equation}
R(\zeta , T)=R(\zeta)\mathrm{e}^{-\rmi\theta T}\label{Revol}
\end{equation}
with
\begin{equation}
\theta = \zeta^2 + \frac{\omega_0}{12}\fint_{-\infty}^{+\infty}\frac{d\lambda}{\lambda - \zeta}\left|A(\lambda)\right |^2 \hskip1mm ,
\end{equation}
where the slashed integral denotes the Cauchy principal value. It worth remarking that $\theta$ has a positive imaginary part $\theta_I=(\pi /12)\omega_0\left|A(\zeta)\right |^2$, which implies thes exponential growth of $R(\zeta +i0, T)$ in time. We shall see that this fact is the very reason of the instability for the system \ref{system}.\\

{\bf Step 3 :} Solve the Riemann-Hilbert problem \ref{riemhil} in which $R(\zeta , T)$ is given in terms of $R(\zeta , 0)$ by means of \ref{volterraR}. The solution is obtained from the following Cauchy-Green integral equation (omitting the $(X,T)$-dependence)
\begin{equation}
\mu(\zeta)=\pmatrix{1\cr 0 }+\frac{1}{2\pi \rmi} \oint_C\frac{d\lambda}{\lambda -\zeta}\sigma_1\hskip1mm \overline{\mu(\bar{\lambda})}R(\lambda)\mathrm{e}^{2\rmi\lambda X}\label{cauchgr}
\end{equation}
where $C$ represents any contour in the upper half plane ($\Im(\zeta)>0$) equivalent to the real axis. Equation \ref{cauchgr} gives the solution $\mu^+(\zeta,X,T)$ while the solution  $\mu^-(\zeta,X,T)$ is obtained out of \ref{riemhil}.\\

{\bf Step 4 :} Finally the solution $(q, a_1, a_2)$ of the system \ref{system} is obtained from the function $\mu$ through
\begin{equation}
\cases{q(X,T)=2\rmi\hskip1mm\bar{\mu}^{(1)}(X,T) \\ a_i(X,T)=A\mu_i^+(\zeta,X,T)\mathrm{e}^{-\rmi\zeta X}  \qquad , \hskip3mm i=1,2 }\label{formsol}
\end{equation}
where $\bar{\mu}^{(1)}(X,T)$ denotes the coefficient of $\zeta^{-1}$ in the Laurent expansion of $\bar{\mu}_2(\zeta,X,T)$.\\

This last step achieves the resolution of the initial/boundary value problem for the system \ref{system}-\ref{cond}.\\

Note that the equations \ref{volterra} and \ref{riemhil} imply 
\begin{equation}
\displaystyle\lim_{X\to +\infty}\mu_2^+=R\hskip1mm \mathrm{e}^{2\rmi\zeta X}\hskip1mm .
\end{equation}
Hence, $R(\zeta +\rmi 0)$ is in fact the recflection coefficient of the potential $q$ at any fixed time $t$. It is easy to verify that the transmission coefficient $\mathcal{T}$, defined by
\begin{equation}
\displaystyle\lim_{X\to -\infty}\mu_1^+ =\mathcal{T} \hskip1mm ,
\end{equation}
and the reflection coefficient $R$ obey the unitarity relation
\begin{equation}
R(\zeta + \rmi 0)\hskip1mm \bar{R}(\zeta - \rmi 0) + \mathcal{T}(\zeta + \rmi 0)\hskip1mm \bar{\mathcal{T}}(\zeta -\rmi 0) = 1\hskip1mm .\label{unit}
\end{equation}

\section{Conservation laws}

\hskip9mm The first usual consequence of the integrability of a system is the existence of an infinite sequence of local conservation laws.\\

The explicite calculation of conservation laws requires the Lax pair $\Omega(k)$ and $W(k)$ \cite{cd}. Here, $\Omega$ is the (singular) dispersion relation (this choice is justified in the Appendix. See \ref{disprel} and \ref{matrixevol})
\begin{equation}
\Omega = \rmi k^2\sigma_3 + \frac{1}{2\rmi\pi}\iint_\mathbb{C}\frac{d\lambda \wedge d\bar{\lambda}}{\lambda -k}g(\lambda)\sigma_3
\end{equation}
and $W$ can be computed with the methode exposed in \cite{lelat}
\begin{equation}
W(k) = -\rmi\sigma_3k^2 + Q\hskip1mm k +\frac{\rmi}{2}\sigma_3\hskip1mm ({Q_{}}_X - Q^2)-\frac{1}{2\rmi\pi}\iint_\mathbb{C}\frac{d\lambda \wedge d\bar{\lambda}}{\lambda -k}g(\lambda)\psi\sigma_3\psi^{-1}
\end{equation}
where $Q = \displaystyle{\pmatrix{ 0 & q \cr \bar{q} & 0 }}$ and $\psi$ is defined in \ref{spect} and \ref{defpsi}.\\

Using the compatibility condition
\begin{equation}
U_T-W_X + \left [ U\hskip 1mm , \hskip1mm W \right ] = 0\hskip1mm ,
\end{equation}
the infinite sequence of conservation laws can be obtained by equating the powers of $\displaystyle{(2ik)^{-1}}$. The first law of this sequence reads
\begin{equation}
\partial_T\left | \hskip1mm q \hskip1mm \right |^2 = \partial_X\left [ \frac{\rmi}{2}\hskip1mm (q_X\bar{q}-\bar{q}_Xq)+\frac{\omega_0}{24}\hskip1mm \mathcal{J} \right ]\label{energy}
\end{equation}
with
\begin{equation}
\mathcal{J} = \int_{-\infty}^{+\infty}d\zeta \left [ a_1(\zeta + \rmi 0)\bar{a}_1(\zeta - \rmi 0) +  a_2 (\zeta + \rmi 0)\bar{a}_2(\zeta - \rmi 0)\right ]\hskip1mm .
\end{equation}
Now by integrating the equation \ref{energy} with respect to $X$ using \ref{cond}, \ref{riemhil} and \ref{unit}, one obtains
\begin{equation}
\frac{\partial}{\partial T}\int_{-\infty}^{+\infty}dX'\hskip1mm \left | q\right | ^2=\int_{-\infty}^{+\infty}d\zeta \left | A'\right | ^2R(\zeta + \rmi 0)\bar{R}(\zeta - \rmi 0)
\end{equation}
where $\displaystyle{ \left | A'\right |=(\omega_0 /12)\left | A\right |}$.\\

Due to \ref{Revol}, the product $R(\zeta + \rmi 0)\bar{R}(\zeta - \rmi 0)$ is time-independent and therefore, the energy $\displaystyle{\int_{-\infty}^{+\infty}dX'\hskip1mm \left | q\right | ^2}$ transferred from the EW to the ISW, grows linearly in time. This causes the instability of the ISW which is going to be described in the next section by evaluating the time-asymptotic solution of $q(X,T)$.

\section{The time asymptotic solution}

\hskip9mm As show in equation \ref{Revol}, the reflection coefficient $R(\zeta+\rmi 0,T)$ has an exponential growth which is characteristic of integral evolutions having a singular dispersion law \cite{leon,kp,man}. Following \cite{man}, the asymptotic behavior as $T\rightarrow +\infty$, of the solution of \ref{cauchgr} is evaluated by the saddle point method.\\

Let consider an arbitrary initial local distribution $q(X,0)$ on the charge density by considering some generic reflection coefficient
\begin{equation}
R(\zeta+\rmi 0\hskip1mm ,\hskip1mm 0)=\frac{r_0}{{(\zeta+\rmi 0})^n}\hskip1mm ,
\end{equation}
where $r_0$ is a real constant and $n > 1$. Let consider also a very sharp (i.e. monochromatic) input laser beam, so that the evolution \ref{Revol} is reduced to
\begin{equation}
R(\lambda)\sim r_0\lambda^{-n}\exp \left [ \rmi T (\frac{\gamma}{\lambda}-2\lambda^2)\right ]\label{simplR}
\end{equation}
for $\Im (\lambda) > 0$ and $\gamma = \displaystyle{\int_{-\infty}^{+\infty}d\zeta ' \left |A'(\zeta ')\right |^2}$. From \ref{simplR}, the saddle point as $T\rightarrow +\infty$, is
\begin{equation}
\lambda_s=\frac{1+\rmi\sqrt {3}}{2}{(\frac{\gamma}{4})^{}}^{(1/3)}\hskip1mm .
\end{equation}
One may choose for the contour $C$ in  \ref{cauchgr}, any contour equivalent to the straight line parallel to the real $\lambda$-axis passing thought this saddle point. This contour, equivalent to the real axis, is precisely the critical path. Then the asymptotic behavior of the solution of \ref{cauchgr} satisfies the following algebric system ($\Im(\zeta)>0$)
\begin{equation}
\pmatrix{\mu_1(\zeta)\cr \mu_2(\zeta) }=\pmatrix{1\cr 0 }+\frac{T^{-(1/2)}}{\lambda_s-\zeta}B\exp \left [ 2\rmi\lambda_sX+\rmi T(\frac{\gamma}{\lambda_s}-2\lambda_s^2) \right ] \pmatrix{\bar{\mu}_2(\bar{\zeta)}\cr \bar{\mu}_1(\bar{\zeta)} },
\end{equation}
where the constant $B$ is 
\begin{equation}
B=\sqrt{\frac{\pi}{6}}\hskip1mm\frac{r_0}{{(\lambda_s)}^{-n}}\hskip1mm\frac{1}{2\rmi\pi}\hskip1mm .
\end{equation}
The solution of the above algebraic system gives the behavior of $\mu^-(\zeta,X,T)$ and $\mu^+(\zeta,X,T)$ through \ref{riemhil} and hence, $a_1$ and $a_2$ through \ref{formsol}. Namely
\begin{equation}
\cases{a_1(X,T) =\displaystyle{ (1-\frac{\bar{\lambda_s}}{\lambda_s})\hskip1mm\frac{\left |\Upsilon(X,T)\right |^2}{1-\left |\Upsilon(X,T)\right |^2}}\\
a_2(X,T) =\displaystyle{(1-\frac{\bar{\lambda_s}}{\lambda_s})\hskip1mm \frac{\Upsilon(X,T)}{1-\left |\Upsilon(X,T)\right |^2}}
}\label{sol1}
\end{equation}
with
\begin{equation}
\Upsilon(X,T) = \frac{\lambda_s}{\lambda_s-\bar{\lambda}_s}\frac{1}{2\rmi\pi}\left [ \frac{-2\pi}{T\omega ''(\lambda_s)} \right ]^{\frac{1}{2}}\alpha(\lambda_s)\hskip1mm\rme^{\displaystyle{\rmi\Phi}}
\end{equation}
where,
\begin{equation}
\Phi=2\rmi\lambda_sX+2\rmi(\lambda_s^2-\frac{\gamma}{2\lambda_s})T 
\end{equation}
and $\omega '' $ stands for the second derivative of $\omega (\lambda)=2\rmi \left [ \displaystyle{\frac{\lambda X}{T} + \lambda^2-\frac{\gamma}{2\lambda}} \right ]$ with respect to $\lambda$.\\

Finally the time asymptotic behavior of $q$ is obtained through \ref{formsol} 
\begin{equation}
\displaystyle\lim_{T\to +\infty}q(X,T)=-2 \Im (\lambda_s) \frac{\mathrm{ e}^{\rmi\phi}}{\sinh \rho}\hskip1mm , \label{sol2}
\end{equation}
with
\begin{equation}
\phi = 2\Re (\lambda_s)\left [ X+ 6 \Re (\lambda_s) T \right ] + \phi_0
\end{equation}
and
\begin{equation}
\rho = 2 \sqrt{3} \hskip1mm \Re (\lambda_s)\left [ X-\Re (\lambda_s)\hskip1mm  T + \ln (\frac{\beta \alpha}{4 \Im (\lambda_s) \sqrt{6\pi}T})\right ]
\end{equation}
where the constants $\alpha$ and $\phi_0$ are related via
\begin{equation}
\frac{r_0}{{(\lambda_s)}^n}=\alpha \hskip1mm\mathrm{e}^{\displaystyle{\rmi\phi_0}} \hskip1mm .
\end{equation}
The consistency of the solution \ref{sol1} and \ref{sol2} can easily be verified by replacing these solutions into the initial system \ref{system} satisfying the initial/boundary condition \ref{cond}.

\section{Conclusions and discussion}

\hskip9mm In conclusion, we make the following comments :\\

\begin{enumerate}
\item Although the asymptotic behavior of $ q(X,T)$ \ref{sol2} is a singular expression,  $q(X,T)$ itself, is not singular. Thus the direct consequence of \ref{sol2} is the growing amplitude of $ q(X,T)$ which is the signature of the energy transfer.
\item The singular point in  \ref{sol2} travels with the asymptotic velocity $-6\hskip1mm \Re (\lambda_s)$. Hence, the solution $q(X,T) $ accumulates energy in the region where the laser beam is applied, which accounts for the low penetration of the laser in the plasma.
\item The basic system \ref{system} is obtained as the small amplitude limit of the fluid type behavior of plasma and the Maxwell equations. Therefore, it is proved here, that this system does not propagate stable small solutions when there is a balance between SBS emission and the ISW nonlinearity. It worth remarking that, by neglecting the ponderomotive effect (which means no right hand side in the evolution equation in \ref{system}), one would obtain an equation for the ISW where any initial disturbance would eventually disperses away as $T\rightarrow \infty$. Therefore, while the nonlinearity in the left hand side of the evolution equation in \ref{system} is a mechanism for the saturation of the (linear) instability of the ISW, the nonlinear ponderomotive effect of the electric field is the mechanism responsible for the plasma instability.
\item It has been admitted that $q(X,T) \in L^1(\mathbb{R})$ which implies in particular that $\left | q(X,T)\right | \rightarrow 0$ as $\left | X \right | \rightarrow \infty$. It is interesting to wonder whether different asymptotic behaviors might change the stability properties of the system. For instance, the large thermal conductivity of the electrons might act as a saturation mechanism.
\item In many physical situations the length of the plasma is finite. Then, arises the question of the properties of the system \ref{system} on a finite X-interval. It is possible, as in the study \cite{clp}, to observe in this case a chaotic behavior.
\item It is well known since the pioneering work of Zakharov \cite{zak}, that the route to Langmuir turbulence passes through the "collapsing of the EW". i.e. the formation of local singularities of the wave amplitude (and hence also of the ISW amplitude named cavitons). This result has been originally obtained in 3 dimension with spherical symmetry and later, in one dimension \cite{pdr} but with higher order nonlinearity having no physical origin. Here we have a model equation, derived from the basic hydrodynamic-Poisson-Maxwell system equations, whose solution generically evolves through a singular solution.
\item Note finally that the monochromatic approximation led to the evolution \ref{simplR} and subsequently to only one saddle point and one singularity in the time-asymptotic behavior of the ISW. Releasing the monochromatic approximation, will introduce more and more saddle points and hence, more and more singular points or asymptotically collapsing waves.
\end{enumerate}

\section*{Appendix : Proof of the method of resolution}
\setcounter{section}{6}

The differential equations
\begin{equation}
\cases{
{a_1}_{,X}+\rmi\zeta a_1=qa_2\\
{a_2}_{,X}-\rmi\zeta a_2 =\overline{q}a_1
}\hskip3mm\displaystyle,\label{extract}
\end{equation}
extracted from the system \ref{system} is the vectoriel form of the Zakharf-Shabat spectral problem \cite{zasha,ksi} that can be written in the general matrix form \cite{kp,leon2,leon3}
\begin{equation}
\psi_x = U \hskip1mm \psi \qquad , \qquad U = -k\hskip1mm \sigma_3 + Q \qquad , \qquad Q = \pmatrix{ 0 & q \cr r & 0 }\label{spect}
\end{equation} 
The case \ref{extract} is recovered by the reduction $r = \bar{q}$, which is going to be adopted from now on. Here $\psi$ is a $2 \times 2$ matrix, built with two independent column vector solutions $(\psi_1\hskip1mm , \psi_2)$ and is completely determined by giving its asymptotic behavior. Then the set of the differential equations $\psi_x = U \hskip1mm \psi$ can be equivalently written as a set of Volterra integral equations. For convenience, let write these equations for the matrix $\mu (k,x)$, defined by
\begin{equation}
\mu = \psi\exp (\rmi k\sigma_3 x)\hskip1mm \Rightarrow \hskip1mm \mu_x =\rmi k\left [ \mu \hskip1mm, \hskip1mm \sigma_3 \right ] + Q\mu\hskip1mm .\label{defpsi}
\end{equation}
Two dependant solutions $\mu^+$ and $\mu^-$ can be defined through
\begin{equation}
\fl
\cases{
\mu_{11}^+=1-\displaystyle{\int}_x^{+\infty}dx'\hskip1mm q\hskip1mm \mu_{21}^+\\
\mu_{21}^+=\displaystyle{\int}_{-\infty}^xdx'\hskip1mm \bar{q}\hskip1mm \mu_{11}^+\hskip1mm \mathrm{e}^{2\rmi k(x-x')}\\
\mu_{12}^+=-\displaystyle{\int}_x^{+\infty}dx'\hskip1mm q\hskip1mm \mu_{22}^+\hskip1mm \mathrm{e}^{-2\rmi k(x-x')}\\
\mu_{22}^+=1-\displaystyle{\int}_x^{+\infty}dx'\hskip1mm \bar{q}\hskip1mm \mu_{12}^+
}\hskip3mm,\\
\cases{
\mu_{11}^-=1-\displaystyle{\int}_x^{+\infty}dx'\hskip1mm q\hskip1mm \mu_{21}^-\\
\mu_{21}^-=\displaystyle{\int}_x^{+\infty}dx'\hskip1mm \bar{q}\hskip1mm \mu_{11}^-\hskip1mm \mathrm{e}^{2\rmi k(x-x')}\\
\mu_{12}^-=-\displaystyle{\int}_{-\infty}^xdx'\hskip1mm q\hskip1mm \mu_{22}^-\hskip1mm \mathrm{e}^{-2\rmi k(x-x')}\\
\mu_{22}^-=1-\displaystyle{\int}_x^{+\infty}dx'\hskip1mm \bar{q}\hskip1mm \mu_{12}^-
}\hskip3mm.\label{gros}
\end{equation}
One can easily verify, by using the Leibnitz formulas, that $\mu$ satisfies \ref{defpsi}. In the reduction $r = \bar{q}$, one has 
\begin{equation}
Q=\sigma_
1\hskip1mm\bar{Q}\hskip1mm\sigma_1 \qquad , \qquad \sigma_1 =\pmatrix{ 0 & 1 \cr 1 & 0 }
\end{equation}
and it is easy to prove that
\begin{equation}
\mu^+(k,x)=\sigma_1\hskip1mm \overline{\mu^-}(\bar{k},x)\hskip1mm \sigma_1\hskip1mm .\label{murel}
\end{equation}
The set of Volterra equations \ref{gros} can be mapped into a Riemann-Hilbert problem as follows :\\

$\mu^+$ (respectively $\mu^-$) is holomorphic in the complex upper half plane $\Im(k)>0$ (respectively in the complex lower half plane $\Im(k)<0$). Let write
\begin{equation}
\mu=\cases{\mu^+ \hskip3mm \textrm{in} \hskip3mm \Im (k)>0\\ \mu^- \hskip3mm \textrm{in} \hskip3mm \Im (k)<0 }\hskip1mm .
\end{equation}
The matrix $\mu$ is holomorphic all over the complex plane except on the real axis where it stands a discontinuity. Writing the Riemann-Hilbert problem consists on expressing this discontinuity in terms of $\mu^+$ (or $\mu^-$). \\

To do so, let Compute
\begin{equation}
D(k,x)=\left [ \mu_1^+(k+\rmi0\hskip1mm, \hskip1mm x) - \mu_1^-(k-\rmi0\hskip1mm, \hskip1mm x)\right ] \mathrm{e}^{2\rmi kx}\qquad , \hskip3mm x\in \mathbb{R}
\end{equation}
out of the integral equations \ref{gros} and obtain the following Volterra equations for $D_1$  and $D_2$, components of $D$.
\begin{equation}
\cases{
D_1=-\displaystyle{\int}^{+\infty}_xdx'\hskip1mm q\hskip1mm D_2\hskip1mm \mathrm{e}^{2\rmi k(x'-x)}\\
D_2=\alpha^+(k)-\displaystyle{\int}^{+\infty}_xdx'\hskip1mm \bar{q}\hskip1mm D_1
}\label{vectdiff}
\end{equation}
where
\begin{equation}
\alpha^+(k)=\int_{-\infty}^{+\infty}dx'\hskip1mm \bar{q}(x')\hskip1mm\mu_{11}^+(k,x')\hskip1mm \mathrm{e}^{-2\rmi kx'}\hskip2mm ,\hskip2mm \Im(k)=0^+\hskip1mm .
\end{equation}
An integral equation having the same Green function as \ref{vectdiff} can be obtained readily out of \ref{gros}. Indeed, the vector
\begin{equation}
L=\mu_2^+(k+\rmi0\hskip1mm ,\hskip1mm x)\hskip1mm \alpha^+(k)\hskip2mm ,\qquad k\in \mathbb{R}
\end{equation}
is also a solution of \ref{vectdiff}. General theorems about integral equations \cite{bookeqdiff} allow to prove that \ref{vectdiff} has only the trivial solution $\alpha =0$. Then we have $D=L$, which gives the following Riemann-Hilbert problem
\begin{equation}
\mu_1^+(k+\rmi0\hskip1mm ,\hskip1mm x) - \mu_1^-(k-\rmi0\hskip1mm ,\hskip1mm x)=\alpha^+(k)\hskip1mm\mathrm{e}^{2\rmi kx}\hskip1mm \mu_2^+(k+\rmi0\hskip1mm ,\hskip1mm x)\hskip1mm .\label{RH1}
\end{equation}
A similar calculation with
\begin{equation}
D'(k,x)=\left [ \mu_2^+(k+\rmi0\hskip1mm, \hskip1mm x) - \mu_2^-(k-\rmi0\hskip1mm, \hskip1mm x)\right ] \mathrm{e}^{-2\rmi kx}\qquad , \hskip2mm x\in \mathbb{R}
\end{equation}
leads to the second Riemann-Hilbert problem
\begin{equation}
\mu_2^+(k+\rmi0\hskip1mm ,\hskip1mm x) - \mu_2^-(k-\rmi0\hskip1mm ,\hskip1mm x)=\alpha^-(k)\hskip1mm\mathrm{e}^{-2\rmi kx}\hskip1mm \mu_2^-(k-\rmi0\hskip1mm ,\hskip1mm x)\hskip1mm .\label{RH2}
\end{equation}
Using \ref{gros}, we have
\begin{equation}
\alpha^-(k)=-\displaystyle{\int}_{-\infty}^{+\infty}dx'\hskip1mm q(x')\hskip1mm \mu_{22}^-(k,x')\hskip1mm \mathrm{e}^{2\rmi kx'}\equiv \-\overline{\alpha^+}(k)\hskip5mm ,\hskip2mm \Im(k)=0^-
\end{equation}
Then \ref{RH1} and \ref{RH2} can be written in the following matrix form
\begin{equation}
\mu^+-\mu^-=(\mu_1^-\hskip1mm ,\hskip1mm \mu_2^+)\hskip1mm S\hskip1mm .\label{RH3}
\end{equation}
with
\begin{equation}
S(k,x)=\mathrm{e}^{-\rmi k\sigma_3s}\hskip1mm \pmatrix{ 0 & -\overline{\alpha^+}(k)\cr \alpha^+(k) & 0}\hskip1mm\mathrm{e}^{\rmi k\sigma_3x}\hskip2mm ,\hskip2mm k\in\mathbb{R}\hskip1mm .\label{Smat}
\end{equation}
Note that , as $\mu$ in \ref{murel}, the matrix  $S$ satisfies the reduction relation
\begin{equation}
S(k,x)=-\sigma_1\hskip1mm \bar{S}(k,x)\hskip1mm \sigma_1\hskip1mm .\label{redmat}
\end{equation}
The function $\alpha^+(k)$ is called the reflection coefficient. $\alpha^+(k)$ and $q(x)$ are equivalent in the sens that $q(x)$ being given, one solves the Volterra equations \ref{gros} and computes $\alpha^+(k)$ through
\begin{equation}
\alpha^+(k)=\displaystyle\lim_{x\rightarrow +\infty}\mu_{21}^+(k,x)\mathrm{e}^{-2\rmi kx}\hskip1mm ,
\end{equation}
which solves the direct problem.\\

The inverse problem consists on constructing $Q(x)$ from a given $S(k,x)$ and solve the Riemann-Hilbert problem \ref{RH3} to get $\mu(k,x)$. Once $\mu$ is obtained, then $q(x)$ is computed in the following way :\\

Write the Laurent expansion
\begin{equation}
\cases{
\mu_{11}^-=1+\displaystyle\frac{1}{k}{\mu_{11}^-}^{(1)}+\displaystyle\frac{1}{k^2}{\mu_{11}^-}^{(2)}+.... \\
\mu_{21}^-=\displaystyle\frac{1}{k}{\mu_{21}^-}^{(1)}+\displaystyle\frac{1}{k^2}{\mu_{21}^-}^{(2)}+ ....
}\label{laurentexp}
\end{equation}
obtained from the Volterra equation \ref{gros} by integrating by parts. Then, insert \ref{laurentexp} in \ref{defpsi} and use the Liouville theorem to obtain
\begin{equation}
q=2\rmi\hskip1mm {\overline{\mu_{21}^-}}^{\hskip1mm (1)} \label{invpb}
\end{equation}
which gives $q(x)$ out of $\mu (k,x)$.\\

To solve the Riemann-Hilbert problem, it is convenient to use the \lq\lq{}DBAR problem\rq\rq{} formulation :
\begin{equation}
\frac{\partial\mu}{\partial\bar{k}}\doteqdot \frac{1}{2}\left [ \mu(k)\hskip1mm \delta^+({k_{}}_I) - \mu(k)\hskip1mm \delta^-({k_{}}_I)\right ]
\end{equation}
where
\begin{equation}
\frac{\partial}{\partial\bar{k}}=\frac{\rmi}{2}(\frac{\partial}{\partial{k_{}}_R}+\rmi\frac{\partial}{\partial{k_{}}_I})\hskip1mm , 
\end{equation}
with
\begin{equation}
k={k_{}}_R+\rmi\hskip1mm {k_{}}_I\hskip4mm , \hskip4mm \delta^+=\delta({k_{}}_I-\rmi0) \hskip4mm , \hskip4mm \delta^-=\delta({k_{}}_I+\rmi0)\hskip1mm ,
\end{equation}
$\delta$ being the usual Dirac distribution.\\

Thus the equation \ref{RH3} becomes
\begin{equation}
\frac{\partial\mu}{\partial\bar{k}}=\mu\hskip1mm R \qquad , \qquad R=\frac{1}{2}\hskip1mm S \hskip1mm \pmatrix{ \delta^+  &  0 \cr 0  &  \delta^- }
\end{equation}
And the generarized Cauchy forrmula reads 
\begin{equation}
\mu (k,x) = \frac{1}{2\rmi\pi}\int_{\partial\mathcal{D}}\frac{d\lambda}{\lambda - k}\mu(\lambda,x)+\frac{1}{2\rmi\pi}\iint_\mathcal{D}\frac{d\lambda \wedge d\bar{\lambda}}{\lambda -k}\mu(\lambda,k)R(\lambda ,k) \label{dbar}
\end{equation}
with $\mathcal{D}$ the complex plane. \\

The Laurent expansion \ref{laurentexp}, allows the computation of the first integral in \ref{dbar} and, taking into account the analytic properties of $\mu$, the second integral of \ref{dbar} is reduced to an integration over the real line. Consequently, for the first column vector $\mu_1$ with the reduction of \ref{murel}, we have
\begin{equation}
\mu_1^-(k,x)=\pmatrix{ 1 \cr 0 } +\frac{1}{2\pi \rmi}\int_{-\infty}^{+\infty}\frac{d\lambda}{\lambda -k}\hskip1mm \sigma_1\hskip1mm \overline{\mu_1^-}(\bar{\lambda},x)\hskip1mm \alpha^+(\lambda)\hskip1mm \mathrm{e}^{2\rmi\lambda x}\hskip1mm .
\end{equation}
A similar equation holds for the second column vector.\\

Hence the inverse problem (construction of $q(x)$ from $\alpha^+(k)$) is resolved by solving the Cauchy-Green integral equation and by calculating $q(x)$ out of \ref{invpb}.\\

Notice that the preceding formalism remains valid if $q(x)$ is assumed to depend also on a real external parameter $t$ (time). Then,  the eigenfunction $\mu(k,x)$ and the spectral transform $R(k,x)$ will also depend on $t$.\\

On the next step, we must construct integral evolutions out of a simple choice of a given time dependence of the spectral transform $R(k,x,t)$. \\

We start with the following generic DBAR problem
\begin{equation}
\cases{
\displaystyle\frac{\partial}{\partial\bar{k}}\hskip1mm \mu (k) = \mu(k)\hskip1mm R(k)\\
\mu (k) = 1 +O(\frac{1}{k})\hskip2mm , \hskip2mm \left | k\right | \rightarrow \infty
}\label{dbarvect}
\end{equation}
and ask for $R$ an $(x,t)$-dependance through
\begin{equation}
\cases{
R_t=\left [ R \hskip1mm , \hskip1mm \Omega \right ]\\
R_x=\left [ R \hskip1mm , \hskip1mm \Lambda \right ]
}
\end{equation}
where $\Lambda$ and $\Omega$ are given distributions of $k\in \mathbb{C}$, functions of $x$ and $t$ \cite{kp,leon2,leon3}.\\

It is a simple task to check the following relations
\begin{equation}
\cases{
\displaystyle\frac{\partial}{\partial\bar{k}}(\mu_x\mu^{-1}-\mu\Lambda\mu^{-1})=-\mu\displaystyle\frac{\partial\Lambda}{\partial\bar{k}}\mu^{-1}\\
\displaystyle\frac{\partial}{\partial\bar{k}}(\mu_t\mu^{-1}-\mu\Omega\mu^{-1})=-\mu\displaystyle\frac{\partial\Omega}{\partial\bar{k}}\mu^{-1}
}\hskip1mm .\label{systemdbar}
\end{equation}
By integrating the above equations, one obtains
\begin{equation}\fl
\cases{
\mu_x(k,x,t)=\left [ U-\displaystyle{\frac{1}{2\pi\rmi}}\displaystyle{\iint}_\mathbb{C}\frac{d\lambda\wedge d\bar{\lambda}}{\lambda -k}\mu(\lambda,x,t)\displaystyle{\frac{\partial\Lambda(\lambda,x,t)}{\partial\bar{\lambda}}} \mu^{-1}(\lambda,x,t)\right ]\times&\\\hskip75mm
\mu(k,x,t)+\mu(k,x,t)\Lambda(k,x,t) \\
\mu_t(k,x,t)=\left [ V-\displaystyle\frac{1}{2\pi \rmi}\displaystyle\iint_\mathbb{C}\frac{d\lambda\wedge d\bar{\lambda}}{\lambda -k}\mu(\lambda,x,t)\displaystyle\frac{\partial\Omega(\lambda,x,t)}{\partial\bar{\lambda}}  \mu^{-1}(\lambda,x,t)\right ] \times&\\\hskip75mm
\mu(k,x,t)+\mu(k,x,t)\Omega(k,x,t)
}\label{grosdbarsys}
\end{equation}
in which the matrices $U$ and $V$ (the "constants" of integration of the $\bar{\partial}$ operator) are given by
\begin{equation}
\cases{
U(k,x,t)=-Pol_k\left [ \mu(k,x,t)\Lambda(k,x,t)\mu^{-1}(k,x,t) \right ]\\
V(k,x,t)=-Pol_k\left [ \mu(k,x,t)\Omega(k,x,t)\mu^{-1}(k,x,t) \right ]
}
\end{equation}
where $Pol_k\left [....\right]$ means "the polynomial part in $k$ of $....$".\\

To obtain the equation \ref{defpsi}, we choose and adopt from now on 
\begin{equation}
\Lambda=\rmi k\hskip1mm \sigma_3\hskip1mm .
\end{equation}
The compatibility condition $\mu_{xt}=\mu_{tx}$ when $\partial\Lambda / \partial\bar{k}=0$, leads to 
\begin{eqnarray}
\fl
U_t(k)-V_x(k)+\left [ U(k) , V(k) \right ] =\nonumber
\\-\displaystyle\frac{1}{2\pi \rmi}\displaystyle\iint_\mathbb{C}\frac{d\lambda\wedge d\bar{\lambda}}{\lambda -k}\hskip1mm \left [  U(\lambda)-U(k)\hskip1mm ,\hskip1mm\mu(\lambda)\displaystyle\frac{\partial\Omega(\lambda)}{\partial\bar{\lambda}}\mu^{-1}(\lambda)\right ] .\label{tevol}
\end{eqnarray}
Now, with the choice
\begin{equation}
\Omega= \rmi k^2\sigma_3 + \frac{1}{2\rmi\pi}\iint_\mathbb{C}\frac{d\lambda \wedge d\bar{\lambda}}{\lambda -k}g(\lambda)\sigma_3\hskip1mm ,\label{disprel}
\end{equation}
the time evolution \ref{tevol} is transformed into
\begin{equation}
Q_t-\frac{\rmi}{2}\sigma_3Q_{xx}+i\sigma_3Q^3=\rmi\hskip1mm\left [ \sigma_3\hskip1mm ,\hskip1mm\frac{1}{2\pi\rmi} \iint_\mathbb{C}d\lambda\wedge d\bar{\lambda}g(\lambda)\mu(\lambda)\sigma_3\mu^{-1}(\lambda) \right ]\label{matrixevol},
\end{equation}
which is the evolution equation in \ref{grosdbarsys}.\\

Therefore, it has been proved that when $Q(x,t)$ evolves according to a nonlinear evolution, its spectral transform $R(k,x,t)$ related to $\alpha^{\pm}(k,t)$ by \ref{Smat} evolves in time according to \ref{systemdbar} with $\Omega (k)$ given in \ref{disprel}.\\

In the reduced case $r=\bar{q}$ we have the relation \ref{redmat}, therefore the explicite time evolution of the reflected coefficient $\alpha =\alpha^+(k)$ is given by
\begin{equation}
\alpha_t({k_{}}_R,t) =2\rmi\hskip1mm \left [ {k_{}}_R^2-\frac{\rmi}{\pi}\iint_\mathbb{C}\frac{d{\lambda_{}}_R d{\lambda_{}}_I}{\lambda -({k_{}}_R-\rmi0)}g(\lambda) \right ]\hskip1mm \alpha({k_{}}_R,t)\hskip1mm .\label{redev}
\end{equation}
The example of the physical moded studied in the section 2 justifies the most intresting following subcase :
\begin{equation}
g(\lambda)=\frac{\rmi\pi}{6}\delta({\lambda_{}}_I)\delta({\lambda_{}}_R-k_0)\left | A({k_{}}_R) \right |\hskip1mm .
\end{equation}
Indeed, in the case \ref{redev} gives after integration
\begin{equation}
\alpha(k,t)=\alpha(k,0)\exp \left [2i\hskip1mm \left ( k^2+\frac{1}{6}\int_{-\infty}^{+\infty}\frac{dk\hskip1mm A(k)}{k_0-k}\right )\hskip1mm t \right ]\hskip1mm .
\end{equation}
Clearly the function $\alpha (k)$ has an essential singularity at $k=k_0$. Hence  $\alpha (k)$ is not defined for $k=k_0$ unless $a(k_0,0)\equiv 0$, which actually required by the system \ref{system}. Indeed, for $q(x,0)$ vanishing at both ends of the x-axis, the consistency requires that $a_1\bar{a}_2$ vanishes too. Therefore, as a consequence of equations \ref{RH2} and \ref{formsol}, $\alpha (k)$ must vanish for $k=k_0$. This condition actually determines the parameter $k_0$ from the initial datum  $q(x,0)$, which is the solution of the equation 
\begin{equation}
\alpha(k_0,0)=\int_{-\infty}^{+\infty}dx'\hskip1mm \bar{q}(x')\hskip1mm \overline{\mu^-_{22}}(k_0,x',0)\hskip1mm \mathrm{e}^{-\rmi k_0x'}=0\hskip1mm .
\end{equation}
Hence, it is a property of the system \ref{system} that the initial value  $q(x,0)$ of the ISW determines the small correction of the wave number of the scattered ESW.

\end{document}